\documentclass[superscriptaddress,showpacs,twocolumn,footinbib,amssymb,aps]{revtex4}

\usepackage{graphicx,dcolumn,bm,amsmath,color,bm}

\newcommand{\eq}[1]{Eq.~(\ref{#1})}
\newcommand{\eqs}[1]{Eqs.~(\ref{#1})}
\newcommand{\fig}[1]{Fig.~\ref{#1}}

\newcommand{\eeq}{ \end{equation} }
\newcommand{\beq}{ \begin{equation} }

\newcommand{\eea}{ \end{align} }
\newcommand{\bea}{ \begin{align} }

\newcommand{\bs}{ {\bf s }}

\newcommand{\bhua}{ {\bf \hat{u}}_{1} }
\newcommand{\bhub}{ {\bf \hat{u}}_{2} }

\newcommand{\bfr}{ {\bf r} }
\newcommand{\bk}{ {\bf k} }
\newcommand{\bz}{ {\bf \hat{z}} }
\newcommand{\bes}{ {\bf s} }

\newcommand{\kbt}{k_{B}T}

\begin{document}

\title{Helical buckling in columnar assemblies of soft discotic mesogens}
\author{L. Morales-Anda and H. H. Wensink}
\email{wensink@lps.u-psud.fr}
\affiliation{Laboratoire de Physique des Solides, Universit\'e Paris-Sud  \& CNRS, UMR 8502, 91405 Orsay, France}

\date{\today}

\begin{abstract}

We investigate the emergence of chiral meso-structures in one-dimensional fluids consisting of stacked discotic particles and demonstrate that helical undulations are generated spontaneously from internal elastic stresses. The stability of these helical conformations arises from an interplay between long-ranged soft repulsions and nanopore confinement which is naturally present in columnar liquid crystals. Using a simple mean-field theory based on microscopic considerations we identify generic scaling expressions for the typical buckling radius and helical pitch as a function of the density and interaction potential of the constituent particles. 
\end{abstract}

\maketitle

\section{Introduction}

Chirality is a key signature of life since many biological molecules owe their specific function to their molecular shape and symmetry. Examples range from molecules possessing at least one carbon atom with four different substituents to more complicated helical molecular assemblies such as the DNA double helix.
Apart from the microscopic domain, chirality is expressed abundantly in the macroscopic world, for example in the left- and right-handed helical conformation of snail shells, the spiral patterns of leaves on a plant stem (phyllotaxis), and other biological and inert structures.
The structure of condensed phases are naturally affected by the chirality of the building blocks.  For example, liquid crystals composed of chiral constituents may form extraordinary structures, ranging from cholesterics, chiral smectics and blue phases \cite{gennes-prost,hiltrop}. The particular meso-structure imparted by chiral correlations on the micro-scale, such as a helical director field in cholesterics or a cubic arrangement of defect lines in blue phases, endow these materials with extraordinary optical and mechanical properties. Important applications of chiral mesostructures reside in domains such as non-linear optics \cite{rodrigues2014}, photonics \cite{shibaev_kopp2003} and stereoselective catalysis \cite{palmer2008}. 

Helical structures may also arise in systems of achiral particles. Illustrious examples are spontaneous chiral symmetric breaking of homopolymers composed of attractive non-chiral monomers \cite{magee2006}, the formation of Bernal spiral clusters in patchy colloidal particles \cite{campbell2005, morgan2013} and the self-assembly of achiral magnetic nanoclusters into helical superstructures  \cite{singh2014}.  Helical meso-structures also emerge from packing colloidal particles into cylindrical tubes at high pressure \cite{snir_kamien, oguz_messina} where transitions between different cluster morphologies may occur upon varying the packing load \cite{lohr_yodh}. In fact, coiled, toroidal and helical architectures seem to be a recurring feature in systems of (block co-)polymers  \cite{sevink_jcp2001,yu_confinedpolymers2008} and other soft matter compounds confined in nanopores \cite{wu_naturematerials2004,alcoutlabi2005}. 

Although the formation of the helical meso-structure occurs by means of spontaneous self-assembly,  there should be no preference for a particular helix sense (left- or right-handed)  if the  constituent molecules are achiral. Spontaneous breaking of the chiral symmetry usually happens at the initial stage of the self-assembly process by some combination of fluctuations (of e.g. thermal nature) favoring one chiral symmetry over the other. The local bias toward a particular chiral symmetry then proliferates to mesoscopic length scales by means of some self-sorting and amplification mechanism which secures the uniform mesoscale symmetry of the assembly \cite{dressel2014,garcia_jacs2012}.

In this work we investigate the possibility of helical meso-structures in columnar assemblies of discotic particles with non-chiral pair interactions. Columnar meso-phases are characterized by a two-dimensional crystalline arrangement of columns, each consisting of a quasi one-dimensional stack of particles with a disordered  internal structure.   In discotic liquid crystals supramolecular chirality manifests itself along the columnar direction and can be imparted by chiral intramolecular interactions \cite{hirschberg2000}, or by non-chiral ones via some type of directionality in the face-to-face interactions (so-called $\pi$-stacking), or a helical organization of the orientational or translation degrees of freedom of the discotic particles along the backbone of the stack \cite{vera_columnar}. 

The focus of this paper is to explore the latter type of chirality, namely the one expressed by an effective helical shape of the columnar stack. To this end we devise a microscopic model and use simple mean-field theory to identify generic expressions for the free energy  associated with weak conformational changes  from a linear stack to a helically buckled assembly of stacked particles subject to a confining external field. 

We subsequently apply the theory to columnar meso-phases of charged discotic colloidal particles, a common building block for many natural \cite{kaolin} and synthetic clays   \cite{Brown98,vanderkooijcolumnar, davidson-overview}.  The dominant particle interactions stem from electrostatic repulsions and the stabilization of columnar and other liquid crystal structures can be rationalized on entropic grounds alone \cite{onsager1949, frenkel_nature88}. Particle density, rather than temperature therefore constitutes the chief thermodynamic parameter in these lyotropic systems. 

In order to establish a link between the one-dimensional fluid and the three-dimensional columnar meso-phase, we construct a self-consistent mean-field theory by invoking a simple Lennard-Jones cell-theory for columnar liquid crystals \cite{lennardjones1937,kramerherzfeld-pre2000,graf1999} and identifying the externally imposed confining potential acting on the stack with an effective internal cell potential imparted by adjacent stacks in the columnar lattice. Expressions for the typical amplitude and pitch of  the helical undulations as well as the corresponding free energy differences between the various buckling modes can then be readily established in relation to the particles density and electro-chemical properties of the individual disks via the inter-disk potential.    We believe that the results of this study provide an essential first step  in assessing collective conformational fluctuations and the possibility of spontaneous chiral symmetry breaking in lyotropic columnar assemblies.

\section{Model}

\begin{figure}
\includegraphics[width=0.8 \columnwidth]{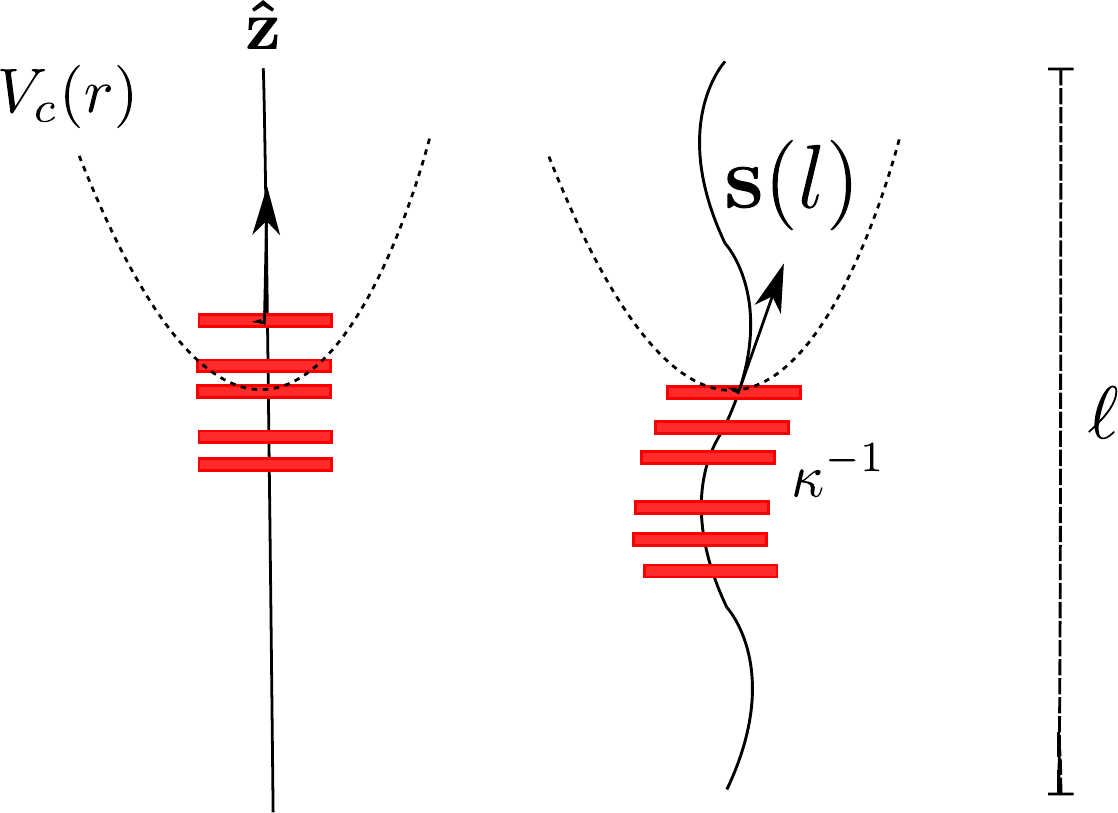}
\caption{A columnar fluid of stacked parallel soft disks (in red) confined in a cylindrical parabolic trap with potential $V_{c}(r)$.  The backbone of the stacks may adopt a linear  (left) or buckled (right) configuration. The typical interaction range between adjacent disks is quantified by a (Debye) screening length $\kappa^{-1}$.}
\label{fig1}
\end{figure}

Let us consider a one-dimensional (1D) fluid of $N$ discotic particles at constant temperature $T$ confined to a cylindrical potential trap $V_{c}(r)$. We focus on dense stacks where the average interparticle distance is very small so that the disks are assumed to be perfectly aligned along the $\bz$ direction of the laboratory frame (see \fig{fig1}).  Their centres-of-mass are constrained to lie on a one-dimensional contour parameterized by $\bes(l) = \{ s_{x}(l), s_{y} (l), s_{z}(l) \} $ with $0 < l < \ell$. The contour length is given by
$ \ell_{c}  = \int_{0}^{\ell} dl  \| \partial_{l} \bes \| $
and  $\ell_{c} > \ell$ for any contour except for a trivial linear one $\bes(l) = \{0,0,l \}$ ($ \ell_{c} = \ell$) which corresponds  to an `unbuckled' system.  Under certain conditions the linearly stacked column may buckle into e.g. a helical shape.

The central idea of this study  is to investigate the optimal buckling shape of the stacked discotic fluid in relation to various control parameters such as particle density, the choice of the pair interaction, and the external potential.  The equilibrium  shape of  the buckled state is not known {\em a priori} but can be established self-consistently using a simple variational approach based on classical density functional theory (DFT). To this end we define the Helmholtz free energy $F$ of a 3D fluid with one-body density $\rho(\bfr)$ defined as the ensemble average $\rho(\bfr) = \langle \hat{\rho}(\bfr) \rangle$ of the microscopic density
$\hat{\rho}(\bfr) =  \sum_{i=1}^{N} \delta (\bfr- \bfr_{i}) $ with normalization $N$. The Helmholtz free energy consists of an ideal, external field and excess contribution, and is formally given by:
\begin{align}
F[\rho]  &= k_{B}T \int d \bfr \rho (\bfr) \left ( \ln \Lambda^{3}  \rho(\bfr)  -1 \right ) \nonumber \\ 
&+ \int d \bfr \rho ( \bfr ) V_{c} (\bfr) + \frac{1}{2}\int d \bfr  \rho(\bfr) \int d \bfr^{\prime} \rho(\bfr^{\prime}) \Phi(  \bfr - \bfr^{\prime} ),
\label{hform}
\end{align}
with $k_{B}T$ the thermal energy ($k_{B}$ denotes Boltzmann's constant) and $\Lambda$ the de Broglie length of each particle. While the first two contributions are exact, the excess term is not. The disk-disk correlations are approximated by a simple {\em mean-field} term featuring the pair potential $\Phi(\bfr)$ between parallel disks. This potential has an intricate directional dependence through the centre-of-mass distance vector $\bfr$ which we will specify later on.  Let us assume the disk centre-of-masses to be distributed uniformly along the strictly one-dimensional contour $\bes(l)$ so that we may  parameterize  $\rho(\bfr) =  \int_{0}^{\ell} d l  \| \partial_{l} \bes \|  \delta (\bfr - \bes(l))$ in terms of a contour-dependent line density $\bar{\rho} = N/\ell_{c}$.  Substituting the parameterization into the free energy yields:
\begin{align}
F  &=   -TS_{\text{id}} + \bar{\rho} \int_{0}^{\ell} d l  \| \partial_{l} \bes \|  V_{c} (\bes (l)) \nonumber \\ 
& + \frac{\bar{\rho}^{2}}{2}\int_{0}^{\ell} d l  \| \partial_{l} \bes \|  \int_{0}^{\ell} d l^{\prime}  \| \partial_{l^{\prime}} \bes \|  \Phi(  \bes(l) - \bes (l^{\prime}) ).
\end{align}
Let us consider a helical parameterization $\bes (l) = \{ R \cos ql, R \sin ql, l \} $ in terms of radius $R$ and pitch $q$. The contour length of a helix is $ \ell_{c} = \ell \sqrt{1+ (qR)^{2}}$. Since a helix has constant curvature, $ \| \partial_{l} \bes \|$ is independent of $l$ and the deformation free energy per particle simplifies to:
\begin{align}
\frac{F}{N}  &=   \frac{-TS_{\text{id}}}{N} +  V_{c} (R) + \frac{\rho}{2}\int_{0}^{\ell} d (l - l^{\prime}) \Phi(  \bes(l) - \bes (l^{\prime}) ),
\end{align}
now featuring a linear density $\rho = N/\ell$ independent of the underlying contour shape. We emphasize that the confining external potential is symmetric in the cylindrical reference frame of the helix and only depends on the lateral extent of the helix via the radius $R$. 
Due to the one-dimensional constraint imposed on the particle contour, the ideal entropy $S_{\text{id}}$ cannot immediately be reproduced from its 3D phase-space definition given by the first contribution in \eq{hform}. In fact, substitution of the delta-distribution for the one-body density into \eq{hform} would lead to a divergence of the free energy.  Nevertheless,  a plausible expression for the ideal entropy compatible with  3D translational motion in real systems can be recovered by factorizing the translational degrees of freedom of each disk into a one-dimensional contribution representing particle motion along the helical contour and a bi-dimensional one originating from infinitesimally small excursions the disk centres-of-mass are allowed to make across the plane normal to the contour.  This will become apparent later on in this text. We will proceed by specifying a suitable form for the inter-disk potential $\Phi$. 

\section{Generalized Yukawa potential for charged disks}

A natural question that arises is what type of pair potential could give rise to stacked fluids with a stable helical meso-structure?  A promising candidate is the effective electrostatic interaction between charged disks of diameter $D$ which in the far-field limit takes the form of an effective Yukawa form \cite{trizac-weis}:
\beq
\Phi (\bfr_{12}, \bhua, \bhub) = u_{0} \xi( \theta_{1} ; \kappa D) \xi(\theta_{2} ; \kappa D) \frac{\exp ( - \kappa r_{12})}{r_{12}},
\label{yukpotential}
\eeq
with amplitude $u_{0} = Z^{2} \lambda_{B}$ in terms of the total number of elementary charges $Z$ per disk, the Debye screening constant  $\kappa$  and $\lambda_{B}$ the Bjerrum length. The angles $\theta$ featuring in the prefactor $\xi$ denote the disk normal orientation relative to the centre-of-mass distance vector $\bfr_{12}$ via  $ \cos \theta = \bz \cdot \hat{\bfr}_{12} $. The angular prefactor reads in explicit form:
\beq
 \xi(\theta; \kappa D)  = 2 \frac{ I_{1} \left (\frac{1}{2} \kappa D \sin \theta \right ) } {  \frac{1}{2} \kappa D \sin \theta  },
 \label{yuk}
 \eeq
with $I_{1}(x)$ a modified Bessel function. The Yukawa potential for discotic colloids favors stacked pair configurations over co-planar ones \cite{trizac-weis}.  As a consequence, the inherently electrostatic anisometry drives self-assembly into stacked meso-phases, including columnar structures \cite{jabbari2013}. Conformational changes from a linear to a buckled state can be anticipated from a trade-off between the external and excess free energy. While the external free energy associated with a confining trap potential increases upon buckling, the cost is offset by a simultaneous reduction of the excess free energy due to the fact that the centres-of-mass between adjacent disks will, on average, be further apart in an undulated stack.  This we will explore in detail in the next Section.

\section{Free energy of a helically deformed  columnar fluid}

Let us now attempt to express the directional Yukawa potential $\Phi$ (\eq{yukpotential}) in terms of the helical vector $\bs(l)$. Since the disks normals are all pointing along the $z$-axis, the only relevant coordinate is the centre-of-mass distance between a pair of particles located at position $l$ and $l^{\prime}$ along the helical contour, i.e., $\Phi(\Delta r)  =\Phi( \bs(l) - \bs(l^{\prime}))$. Without loss of generality, the centre-of-mass distance ${\bf \Delta r} = {\bf r}^{\prime} - {\bf r} $ between two particles located on a helical contour can be expressed in the cylindrical coordinate frame of one of the particles. Straightforward algebra then leads to:
\beq
{\bf \Delta r} = R (\cos (q \Delta l)  -1) {\bf \hat{r}}  + R \sin (q \Delta l )  \hat{\bm \varphi}  + \Delta l \bz,
\label{delr}
\eeq
with $\Delta l = l^{\prime}  - l$. In order to simplify the orientation-dependent prefactor of the Yukawa potential we Taylor expand the Bessel function  \eq{yuk} for small $\theta$. Some manipulation then leads to:
\beq
\Phi (\Delta \bes) =  \tilde{u}_{0}(\cos \theta) \frac{\exp ( - \kappa | \Delta \bfr | )}{| \Delta \bfr |},
\label{phipar}
\eeq
where the amplitude depends on the angle $\theta$ between the centre-of-mass distance and the disk orientation unit vectors aligned along helix axis:
\beq
\tilde{u}_{0}(\cos \theta) = u_{0} \left(  1+  \frac{(\kappa D)^{2}}{16} (1 - \cos \theta) \right )^{2}, 
\eeq
Up to leading order in $q$ the dot product $\cos \theta$ between the centre-of-mass distance and disk orientations can be written as:
 \begin{align}
 \cos \theta & = \frac{| s_{z}(l) - s_{z}(l^{\prime}) | }{\Delta r} \approx \frac{1}{\varepsilon},
 \label{cost}
 \end{align}
in terms of the contour length $\varepsilon = \sqrt{1 + (qR)^{2}}$. The amplitude then becomes:
\begin{align}
\tilde{u}_{0} (\varepsilon)  = u_{0} \left( 1+  \frac{(\kappa D)^{2}}{16} (1 -  \varepsilon^{-1}) \right )^{2}. 
\label{u0}
\end{align}
The effective Yukawa amplitude increases with $q$ and $R$ and thus clearly disfavors buckling. With this the deformation free energy takes the following form:
\begin{align}
\frac{F}{N} &=   \frac{-TS_{\text{id}}}{N} +  V_{c}(R) + \frac{\rho }{2} \tilde{u}_{0} (\varepsilon )  \int_{0}^{\ell}   d \Delta l \frac{\exp [ - \kappa \Delta r  ]  }{ \Delta r },
\end{align} 
with the inter-disk distance $\Delta r$ depending implicitly on the contour parameters $\Delta l$ via \eq{delr}. The last term confronts us with a double contour integral that does not converge for unbounded potentials since the disk centre-of-masses are forced to reside on a one-dimensional contour. The divergence can be lifted by allowing the disks to perform small spatial fluctuations $\delta \bfr_{\perp}$ in the $\left \{ {\bf \hat{r}}, {\hat{\bm \varphi}} \right \} $-plane perpendicular to the main helical direction $\bz$.  The centre-of-mass distance then takes the form:
\beq
{\bf \Delta r} =  \delta {\bf r}_{\perp} + R (\cos (q \Delta l ) -1) {\bf \hat{r}}  + R \sin ( q \Delta l )  \hat{\bm \varphi}  +  \Delta l \bz,
\label{delr_fluc}
\eeq
and the contour integral can then be converted to a three-dimensional via a straightforward weighted average:
\beq
\frac{\rho}{2}\tilde{u}_{0} (\varepsilon)  \int d \delta \bfr_{\perp}  f_{G} (\delta \bfr_{\perp}) \int_{0}^{\ell}  d \Delta l \frac{\exp [ -  \kappa \Delta r  ] }{\Delta r}.
\eeq
The transverse fluctuations obey a 2D-Gaussian distribution with variational parameter $\alpha$:
\beq
f_{G} (\delta r_{\perp}) =  \left ( \frac{\alpha}{\pi} \right ) e^{-\alpha  \delta r_{\perp}^{2}}.
\eeq
In principle, these fluctuations will also have an effect on the  angular contribution $\cos \theta$. However, their impact is expected to be less critical than on the centrally symmetric part and we will retain the form \eq{u0} for simplicity. It is expedient to recast the  expression in  Fourier space in terms of the 2D wavevector $\bk $. This allows us to express the integral in factorized form: 
\begin{align}
 & \frac{\rho}{2} \tilde{u}_{0}(\varepsilon) \int d \delta \bfr_{\perp}  \int \frac{d \bk }{(2 \pi)^{3}}  f_{G} (\delta \bfr_{\perp}) e^{-i \bk  \cdot \delta \bfr_{\perp}}   \hat{U}(k) P(\bk ), 
\end{align}
where $ \hat{U}( \bk  ) =  4 \pi / ( k^{2} + \kappa^{2}) $ is the FT of the spherically symmetric Yukawa potential and $P(\bk)$ a shape factor accounting for the helical shape of the buckled fluid. The Fourier transform only acts on the Gaussian displacement distribution so that:
\begin{align}
 \frac{\rho}{2} \tilde{u}_{0}(\varepsilon) \int \frac{d \bk }{(2 \pi)^{3}}  \hat{f}_{G}(k )  \hat{U}(k) P(\bk ), 
 \label{ftex}
 \end{align}
where the FT of a Gaussian is itself a Gaussian, namely $\hat{f}_{G}(k_{\perp}) = e^{-k_{\perp}^{2}/4\alpha}$.  
The shape factor is formally given by a FT of the helical backbone parameterized in \eq{delr_fluc}:
\beq
P(\bk ) =  \int_{-\ell/2}^{\ell/2}  d \Delta l  \exp [i \bk  \cdot ({\bf \Delta r}  - {\bf \delta r}_{\perp})], 
\label{form}
\eeq
and depends implicitly on the helix shape via $\Delta r$. 

Reverting to the ideal entropy in \eq{hform}, as of yet unspecified, we may now associate the transversal Gaussian fluctuations with an entropic term of the following form:
\begin{align}
 S^{(G)}_{\text{id}}(\alpha)  &\sim -Nk_{B} \int d\delta \bfr_{\perp}  f_{G}(\delta r_{\perp})  [  \ln  \Lambda^{2} f_{G} (\delta r_{\perp})  -1  ]  \nonumber \\ 
 &\sim  -Nk_{B} (\ln(\alpha \Lambda^{2} /\pi) - 2),
\end{align}
indicating a reduction of ideal entropy upon increased localization along the contour. As anticipated,  $S_{\text{id}}$ diverges if the particle centres-of-mass follow a strictly 1D  contour in the limit $\alpha \rightarrow \infty$.
The two-dimensional contribution needs to be supplemented with the entropy arising from the one-dimensional degrees of translational freedom along the contour. This leads to a contribution  $-Nk_{B}( \ln \Lambda \rho  - 1)$ so that the total entropic contribution to the free energy takes the following approximate form:
\beq
S_{\text{id}}  \sim -Nk_{B}   (\ln(\bar{\rho} \alpha \Lambda^{3} /\pi) - 3).
\eeq 
The relevant density variable  $\bar{\rho} = \rho /\sqrt{1+ (qR)^{2}}$ reflects the notion that the available one-dimensional volume should increase upon buckling. In the ideal gas limit ($\Phi \downarrow 0$, $V_{c}  \downarrow 0$)  a linearly stacked system will trivially maximize its configurational entropy by reducing its transverse localization ($\alpha$ decreases) and maximizing its fictitious contour length (although the shape need no longer be a simple helix). 

\eq{form} cannot be worked out in analytical form for arbitrary $R$. Instead we will resort to a series expansion up to quadratic order in $R$ which should be of sufficient accuracy for weakly buckled fluids. The details of this calculation are shown in the Appendix. The generic expression for the deformation free energy per particle reads:
\begin{align}
\frac{F}{N} &\sim   \ln (\rho \Lambda^{3}  \alpha / \pi) - 3  - \frac{1}{2}(qR)^{2}  +  V_{c}(R) \nonumber \\ 
& + \frac{\rho }{2} \tilde{u}_{0} (\varepsilon )   {\mathcal U}(R,q), 
\label{wtot}
 \end{align}
For notational brevity we shall implicitly use  the thermal energy $k_{B}T$ as  energy unit  and $\kappa^{-1}$ as  length unit.   An approximate expression for the interaction free energy  ${\mathcal U}$ of a weakly buckled helical fluid in the strong localization limit  ($\alpha \gg 1$) is derived in the Appendix and reads: 
\begin{align}
{\mathcal U}(R,q) & = \int \frac{d \bk }{(2 \pi)^{3}}  \hat{f}_{G}(k )  \hat{U}(k) P(\bk )  \nonumber \\
& \sim (\pi \alpha)^{1/2} \left ( 1 + R^{2}  \alpha  p_{2} (q \ell) + {\mathcal O} ( (\kappa R)^{4}) \right ),
\label{uquad}
\end{align}
up to quadratic order in $R$. At this point it is instructive to broaden our scope a little by making a distinction between lyotropic systems where Yukawa interactions are repulsive, and  thermotropic ones where an {\em attractive} Yukawa potential could be used as a rough approximation for the  ($\pi$-)stacking or hydrogen-bonding forces between discotic molecules \cite{brandt_michels1995,delrio2005}. In the lyotopic case $\rho u_{0} > 0$ the density is the main thermodynamic variable, whereas 
 for thermotropics  ($\rho u_{0} <0 $)  the potential amplitude plays the role of an effective temperature  $T_{\rm eff}  = | \rho u_{0} |^{-1} $.  From \eq{wtot} and \eq{uquad} we infer that an unbuckled fluid of repulsive disks can reduce its excess  free energy by decreasing $\alpha$ and delocalizing from the contour, while attractive disks tends to favor increased localization along the contour with $F$ growing more negative with $\alpha$. As a consequence,  the buckling scenario for thermotropics will be quite different from that of lyotropics as we will see shortly. 

Contrary to the ideal contribution, which increases with $\alpha$ due to a loss of configurational entropy, the confinement free energy must be inversely proportional with particle localization. The free energy penalty associated with disks moving away from the helical backbone  can be made explicit by considering a parabolic form $V_{c}(r) = v_{0}r^{2}$ with $v_{0} > 0$. The transverse delocalization enters via a simple Gaussian average:
\begin{align}
V_{c}(R, \alpha )  &= v_{0} \int_{0}^{\infty} \frac{ d \delta  r_{\perp}^{2}}{2} \int_{0}^{2 \pi} d \varphi  f_{G}( \delta r_{\perp} )  (R +  \delta r_{\perp} \cos \varphi )^{2}  \nonumber \\
& = v_{0}  \left ( R^{2} + \frac{1}{2 \alpha } \right  ).
\label{vext}
\end{align}
and is inversely proportional to $\alpha$.
In the weak buckling limit ($R \ll 1$) we assume the equilibrium value of $\alpha$ to be unaffected by the undulation so that  $\left . \partial F  / \partial \alpha \right |_{R=0} =0 $. For the lyotropic case $\rho u_{0} >0 $ an approximate analytical solution of the minimization condition can be found in the limit of strong confinement $v_{0} \gg 1$ in which case the equilibrium localization strength is governed by a trade-off  between the external and excess free energy. The extremum value reads:
\begin{align}
\alpha \sim \left (  \frac{2v_{0}}{ \pi^{1/2} \rho u_{0}} \right )^{2/3} \hspace{0.2cm} \rho u_{0} > 0,
\label{extrema_lyo}
\end{align}
showing that the disks become progressively delocalized  with increasing particle concentration. For the thermotropic case the linear fluid is naturally stabilized by the strongly attractive face-to-face interactions and the confining potential will only have a marginal effect on the localization strength. Irrespective of $v_{0} >0 $, the minimal free energy is obtained  in the limit $ \alpha \rightarrow \infty $. In order to systematically assess the buckling free energy in powers of  $R$ we expand the  curvature-dependent Yukawa amplitude \eq{u0} up to quartic order in $qR \ll 1$:
 \beq
\frac{\tilde{u}_{0}( \varepsilon )}{u_{0}}  \approx 1 +  c_{2}^{(L)}(qR)^{2} + c_{4}^{(L)}(qR)^{4}  + {\mathcal O}(R^{6}),
 \eeq
 with  $c_{2}^{(L)} = D^{2}/16$ and $c_{4}^{(L)} =  (D^{4} - 48 D^{2})/1024$. 
It is obvious that the orientation-dependent Yukawa amplitude in \eq{yuk} only makes sense in the context of charge-stabilized colloidal interactions where the Debye screening length is clearly defined.  For the thermotropic case the prefactors  $c_{n}$ must have a different origin as their amplitude and sign will depend on whether face-to-face configurations are energetically favorable compared to edge-to-edge ones or vice versa.   

Going back to  the lyotropic case we may now formulate the free energy difference between the buckled and the unbuckled columnar fluid of up to quartic order in the buckling amplitude:
\begin{align}
 \frac{\Delta F(q)}{NR^{2}} & \sim   - \frac{1}{2}q ^{2} + \frac{1}{4}R^{2}q^{4} + v_{0} + \frac{\rho u_{0}}{2} (\pi \alpha)^{1/2}   \nonumber \\ 
 & \times \left (c_{2}^{(L)} q^{2} +    \alpha p_{2} (q \ell)  +  c_{4}^{(L)} R^{2} q^{4} + \alpha^{2} R^{2} p_{4}(q \ell) \right ).  
 \label{flyo}  
\end{align}
In the limit of strong confinement $\alpha \gg 1$ the free energy is fully dominated by the external and excess free energy. 
Minimization with respect to $R$ yields the equilibrium value for the buckling radius $R^{\ast}$:
\beq
R^{\ast} \propto f(q \ell) \left ( \frac{\rho u_{0}}{v_{0}} \right )^{1/3}  \ll 1,
 \label{rmin}  
\eeq
in terms of the pitch-dependent function $ f(q \ell) = \sqrt{-2 (1 + p_{2}(q \ell) )/ p_{4}(q \ell)}$ (see \eq{p2n} in the Appendix). The $q$ values that minimize the free energy turn out to be independent of the parameter combination $\rho u_{0}/v_{0}$ and follow from $f^{\prime}(q^{\ast} \ell) =0$. The corresponding buckling free energy  is uniformly negative:
\beq
\frac{\Delta F(q^{\ast})}{N} \propto -\frac{v_{0}^{5/3}}{(\rho u_{0})^{2/3}} \frac{(1+ p_{2}(q^{\ast} \ell)) )^{2}}{p_{4}(q^{\ast} \ell) } < 0,
\eeq
 indicating that helical buckling always leads to a reduction of the free energy. On the contrary, for thermotropic systems buckling is generally disfavored. Up to lowest order in radius
  it follows that $ \Delta F(q^{\ast})/NR^{2} \propto \alpha^{3/2} |p_{2} (q^{\ast} \ell)|/ 2 T_{{\rm eff}} \gg 0 $.  This outcome is, of course, not surprising given that the attractive stacking interactions prevent any deformation of the linear stack at low temperatures.

\begin{figure}
\includegraphics[width=0.9 \columnwidth]{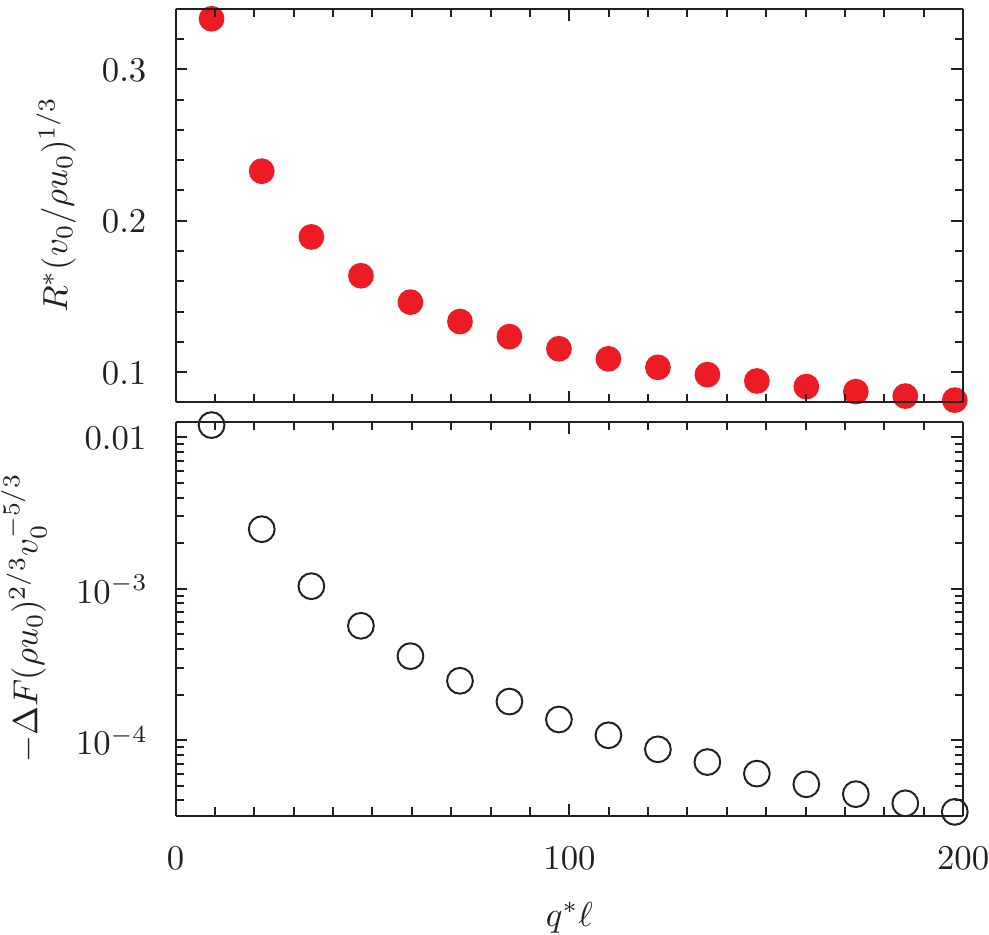}
\caption{Renormalized buckling radius and associated deformation free energy  (on a log scale) for a linear stack of soft discotic particles as a function of the optimal pitch values $q^{\ast} \ell$ . The energetically most favourable mode is $q^{\ast}\ell \approx 9.132$, with corresponding radius $R^{\ast} \approx  0.333 ( \rho u_{0}/v_{0} )^{1/3} $. Higher order  pitch `harmonics' correspond to increasingly weaker buckling radii $R^{\ast}$ and free energy reductions. } 
\label{fig3}
\end{figure}

\section{Connection to lyotropic columnar assemblies}

As an alternative to the parabolic trap we consider a confining potential of the Yukawa form:
\beq
V_{Y}(x) = u_{0} \left ( \frac{1}{2} \frac{ e^{ -\kappa \left ( \sigma- x \right ) }}{\sigma -x }  + \frac{1}{2} \frac{  e^{-\kappa \left ( \sigma + x \right ) } }{\sigma + x}   - \frac{e^{-\kappa \sigma}}{\sigma} \right ),  \hspace{0.2cm} |x| < \sigma,
\eeq
which can be thought of as a simple effective potential acting on each particle along the backbone imparted by its neighbouring columns at a typical distance $ \sigma $ from the target particle. The last term provides the correct offset so that $V_{Y}(0) = 0$. Contrary to the parabolic form, the Yukawa potential thus features a typical interaction range $\sigma$ whose value can either be varied as an independent parameter (cf. the spatial extent of an optical trap), or it can be coupled self-consistently to the thermodynamic state of the columnar assembly. Analogous to \eq{vext} the potential is modified by the  transverse displacements perpendicular to the column direction. 
For small radii an expansion up to quadratic order is appropriate so that:
\beq 
V_{Y}(R, \alpha) = u_{Y} \left ( R^{2} + \frac{1}{2 \alpha}  \right ) + {\mathcal  O} \left ( \frac{R^{4}}{\alpha} \right ).
\label{vcell}
\eeq
We may now simply repeat the steps outlined in the previous section. The localization parameter $\alpha$ still obeys \eq{extrema_lyo} but with $v_{0}$ replaced by an effective value associated with the cylindrical Yukawa potential (denoted ``Y''), namely 
$u_{Y} = u_{0} \sigma^{-3} e^{-\sigma }(1 + \sigma + \sigma^{2}/2)$. In line with the previous section we use $\kappa^{-1}$ as an implicit length unit.

A connection with a three-dimensional columnar phase can be made by coupling the one-dimensional free energy discussed thus far to a simple cell theoretical expression for a (hexagonal) packing of columns. Lennard-Jones cell theory \cite{lennardjones1937} tells us that, approximately:
\beq
Q_{\rm cell}(N) \approx \left ( \int \frac{d^{2} \bfr}{\Lambda^{2}} \exp  \left [- \frac{ V_{Y}(\bfr)}{ 2 \kbt } \right ]  \right )^{N}.
\eeq
Within this framework the line fluids are localized in $N$ cylindrical cells centered on the sites of a fully occupied lattice of some prescribed symmetry. Each particle experiences a potential energy $V_{Y}(R)$ generated by its nearest neighbors. The term between brackets is therefore equivalent to the lateral  `free volume'  each particle explores within its cell.
In its simplest version, the theory presupposes each cell to contain only one stacked fluid behaving independently from its neighbors \cite{lennardjones1937,kramerherzfeld-pre2000,graf1999}. Moreover, the free volume term does not reflect the precise nature of the 2D Bravais lattice of the columnar phase which we assume to be a simple hexagonal arrangement.   Using \eq{vcell} and some algebra then yields for the partition sum:
\beq
Q_{\rm cell}(N) \approx \left ( \frac{\pi   e^{-\frac{u_{Y}}{2 \alpha}} \left ( 1 - e^{-\sigma^{2} u_{Y}} \right ) }{u_{Y} \Lambda^{2}}   \right )^{N},
\label{qcell}
\eeq
Single occupancy of each cell implies that the three-dimensional density $\varrho = N/V$ is linked to the linear density via $ \varrho = \rho /\sigma^{2} $.
Recalling that $F = -\kbt \ln Q$  we obtain the free energy per particle of the columnar phase by superimposing the contributions from the stacked fluid and the cellular confinement (ignoring irrelevant constants) \cite{kramerherzfeld-pre2000}\footnote{The external field term featuring in the fluid free energy must be omitted given that the effect of confinement is accounted for by the cell free energy \eq{qcell}. }: 
\begin{align}
\frac{F(\varrho)}{N }  &\sim  \ln (\varrho \sigma^{2}  \alpha )  + \frac{u_{Y}}{2 \alpha} + \frac{\varrho \sigma^{2}}{2} u_{0}  (\pi \alpha)^{1/2} + \ln u_{Y} \nonumber \\ 
& - \ln    \left (1 - e^{-\sigma^{2} u_{Y}} \right ).   
\label{wcol}
 \end{align}
 In the limit of very strong confinement we expect $\sigma \ll 1$ so that $u_{Y} \sim u_{0} /\sigma^{3} \gg 1$. Minimization of the columnar free energy with respect to $\alpha$ yields, as expected, an inverse proportionality of the localization strength with density and amplitude, i.e.  $\alpha  \sim (4/\pi)^{1/3} \varrho^{-2/3} \sigma^{-10/3}  $. Subsequent minimization of the free energy with respect to $\sigma$ gives a similar proportionality but with different exponents, namely $ \sigma \sim \sigma_{0} /  u_{0}^{3} \varrho^{2} $
with constant $\sigma_{0} = 70304/27\pi$. With this we can establish an explicit relation between the density $\varrho$ of the columnar phase and the localization strength, namely:
\beq
\alpha \sim \alpha_{0} u_{0}^{10} \varrho^{6},
\eeq
with prefactor  $\alpha_{0} = (4/\pi)^{1/3} \sigma_{0}^{-10/3}$. Similarly,  the effective amplitude of the self-consistent cell potential reads:
\beq
u_{Y} \sim \frac{u_{0}^{10} \varrho^{6}}{\sigma_{0}^{3}}. 
\eeq
In order to get a feeling of the magnitude of the variables above we may identify $u_{0} \approx Z^{2} \kappa \lambda_{B} $ with $Z \gg 1$  the number of bare or effective elementary charges per disk \cite{rowan_mp2000,trizac-weis} and $\lambda_{B}$ the Bjerrum length (which is typically about 1 nm in aqueous conditions at room temperature).  The effective  columnar packing fraction can be estimated from $\phi \sim N D^{2} \kappa^{-1} /V $ so that  $\varrho \sim \phi (\kappa^{-1}/D)^{2}<1$. 
The typical buckling radius then scales as:  
\beq
R^{\ast} \propto  \frac{ f(q^{\ast} \ell)}{\varrho^{3} u_{0}^{5}} \ll 1,
 \label{rmincol}  
\eeq
and the deformation free energy associated with the pitch optima (see \fig{fig3}) scales as follows:
\beq
\frac{\Delta F(q^{\ast})}{N} \propto - u_{0}^{20} \varrho^{12}  \frac{(1+ p_{2}(q^{\ast} \ell)) )^{2}}{p_{4}(q^{\ast} \ell) } \ll 0.
\label{minfree}
\eeq
For infinitesimally weak buckling deformations,  a Taylor expansion of \eq{flyo}  up to  the lowest order contribution in the helix parameters gives an expression which could be interpreted as the typical bend or twist elastic modulus of a single column: 
\beq
 \frac{\Delta F}{N(qR)^{2}} \propto - \frac{u_{0}^{10} \varrho^{6} \ell^{2} }{24 \sigma_{0}^{3}}, 
\eeq
The modulus is negative because internal stresses along the backbone  lead to buckling instabilities and  `floppy' modes whose amplitude increases quadratically with length $\ell$. 
We finally wish to point out that the buckling radius and free energy depend implicitly on the Debye screening length. Rearranging the scaling expressions \eqs{rmincol} and (\ref{minfree})  in terms of the (fixed) disk diameter $D$ reveals that $R^{\ast}/D \propto (D/\kappa^{-1})^{3} $ and $\Delta F/ N \propto (\kappa^{-1}/D)^{16}$ and shows that the buckling characteristics depend sensitively on the range of the stacking potential.

\section{Concluding remarks}

We have proposed a microscopic theory to probe helical buckling deformations of a columnar fluid of discotic particles.  Contrary to the more commonly studied case of buckling under an external load, the stresses that stabilize helical confirmations of the columnar backbone are purely internal and are transmitted through a strongly repulsive stacking potential between adjacent disks located along the backbone. The approach is particularly suited to investigate conformational fluctuations in lyotropic columnar assemblies of e.g. clay platelets at low salt concentrations where the effective charge-mediated particle interactions are strongly repulsive. Using a simple mean-field theory based on microscopic principles we derive scaling relations for the optimal pitch, which sets the typical buckling length scale, and amplitude in relation to the density of the columnar mesophase and the strength of the electrostatic interactions.  Our findings indicate that helical deformations of stacked fluids can be stabilized by long-ranged soft interactions.

Our analysis is based on the assumption that the columnar deformations are independent from each other.  However, we consider a detailed knowledge of the single-column buckling behaviour as an essential precursor to tackling the more complicated problem of collective conformational fluctuations.  In this context, we plan to extend the present model such as to include chiral intercolumnar couplings, e.g., by introducing a weakly chiral self-consistent cell potential,  in  an effort to understand if and how helical conformational fluctuations of an individual column proliferate to larger length scales. This could open a route to studying spontaneous chiral symmetry breaking and chiral amplification in columnar assemblies of achiral discotic particles under the influence of internal stresses generated by repulsive long-ranged stacking potentials. The results of the current model are expected to facilitate the construction of coarse-grained models for studying the chiral intra-columnar meso-structure of columnar liquid crystals. These models  could provide a useful intermediate route between fully particle-resolved simulations \cite{duncanPRE2009,martinezhayaPRE2010} and phenomenological continuum theories that 
have been fruitfully employed to study columnar bundles of helical fibres \cite{grason_bruinsma2007} and defects in chiral columnar phases \cite{kamien_nelsonPRE1996}. Research efforts along these lines are currently pursued.

\appendix

\section{Excess free energy of helically stacked soft disks}

Let us decompose the wavevector in a 2D transversal and longitudinal component so that $\int d \bk  = \int d \psi  \int dk_{\perp} k_{\perp} \int d k_{\parallel}$ and $k^{2}  = k_{\perp}^{2} + k_{\parallel}^{2}$. The Fourier integral  \eq{ftex} can then be expressed as follows: 
\begin{align}
 {\mathcal U}(R,q) = \int_{0}^{\infty} \frac{dk_{\perp}}{2 \pi}  k_{\perp}  \hat{f}_{G}(k_{\perp}) \int \frac{dk_{\parallel}}{2 \pi} \hat{U}(k) P_{\psi}(k),
 \label{fint}
\end{align}
in terms of the the angle-averaged shape factor 
\begin{align}
P_{\psi}(k) &= \int_{0}^{2 \pi} \frac{d \psi}{2 \pi}  \int_{-\ell/2}^{\ell/2}  d \Delta l  e^{ i  R\bk_{\perp} \cdot [ (\cos (q \Delta l ) -1) {\bf \hat{r}}  +  \sin ( q \Delta l )  \hat{\bm \varphi} ] } \nonumber \\
 & \times e^{ i k_{\parallel} \Delta l \bz}.
\end{align}
The above integral can be greatly simplified by pre-averaging the pair potential over the longitudinal wave-vectors, an approximation that does not lead to qualitative error \footnote{The difference between the pre-averaged  and numerically exact integrals is found to be a couple of percent at most, irrespective of the pitch.}: 
\begin{align}
 {\mathcal U}(R,q) & \approx \int_{0}^{\infty} \frac{dk_{\perp}}{2 \pi}  k_{\perp}  \hat{f}_{G}(k_{\perp})  P_{\psi}(k_{\perp}) \langle {\hat{U}}(k) \rangle_{\parallel}.
\end{align}
Taking the FT of the Yukawa potential $\hat{U}(k) = 4 \pi / (k_{\perp}^{2}  + \kappa^{2})$ gives: 
\begin{align}
 \langle {\hat{U}}(k) \rangle_{\parallel} & = \int_{-\infty}^{\infty} \frac{dk_{\parallel}}{2 \pi}  \hat{U}(k)   = 2 \pi  / \sqrt{k_{\perp}^{2} + \kappa^{2}},
\end{align}
which now depends only on the transverse Fourier component.  In the weak buckling limit ($R \ll 1$) we may expand $P_{\psi}$ up to quartic order in the buckling amplitude $R$. Retaining only the real-valued contributions yields:
\begin{align}
P_{\psi}(k_{\perp})  &=   1+ \frac{(k_{\perp} R)^{2}}{2} p_{2}(q\ell) + \frac{(k_{\perp} R)^{4}}{32} p_{4} (q \ell)+  {\mathcal O}(R^{6}),
\label{ppsi}
\end{align}
in terms of the following pitch-dependent functions:
\begin{align}
p_{2}(q \ell ) &=   j_{0}  ( q \ell /2  ) -1 \nonumber \\
p_{4} (q \ell) &=  3 - 4 j_{0}(q \ell /2) + j_{0} (q \ell),  
\label{p2n}
\end{align}
with  $j_{0}(x) = \sin x / x$  a spherical Bessel function. Note that  the helix form factor is an even function of $q$  and does not discriminate between a right-handed ($q>0$) or left-handed ($q<0$) helix sense.  
The integrations over the transverse wavenumber $k_{\perp}$ can be performed analytically. In the limit of asymptotically large $\alpha$ it suffices to retain only the leading order powers in $\alpha$:
\begin{align}
\int_{0}^{\infty} \frac{dk_{\perp}}{2 \pi} k_{\perp} \langle \hat{U}(k) \rangle_{\parallel} e^{-\frac{k_{\perp}^{2}}{4 \alpha}} k_{\perp}^{2n} \sim c_{2n} (\alpha),
\label{perpi}
\end{align}
with $c_{0} =( \pi \alpha )^{1/2}$, $c_{2} = 2 \pi^{1/2} \alpha^{3/2}$ and $c_{4} = 12 \pi^{1/2} \alpha^{5/2} $. Combining terms we obtain the following expression for the interactions free energy in the strong localization limit for arbitrary order in the pitch $q$:  
\begin{align}
\frac{ {\mathcal U}(R,q)}{(\pi \alpha)^{1/2}} \sim 1 +  R^{2}  \alpha  p_{2} (q \ell) +  R^{4} \alpha^{2} p_{4} (q \ell) +  {\mathcal O}(R^{6}),
\label{uexp}
\end{align}
with $R$ and $\alpha$ implicitly rendered dimensionless in units $\kappa^{2}$.

\acknowledgments
LMA gratefully acknowledges the CONACyT (Mexico)  for financial support.

\bibliography{pub}

\end{document}